\documentclass[runningheads]{llncs}
\usepackage[T1]{fontenc}
\usepackage{graphicx}
\usepackage{amsmath}    
\usepackage{amssymb}    
\usepackage{booktabs}    
\usepackage{caption}     
\usepackage{multirow}
\begin{document}
\title{CEMG: Collaborative-Enhanced Multimodal Generative Recommendation}
%
%
%
\author{
Yuzhen Lin\inst{1} \and
Hongyi Chen\inst{2} \and
Xuanjing Chen\inst{3} \and
Shaowen Wang\inst{4} \and
Ivonne Xu\inst{5} \and
Dongming Jiang\inst{6}\thanks{Corresponding author: dj37@rice.edu}
}

\institute{
School of Information Systems and Management, Carnegie Mellon University, Pittsburgh, PA 15213, USA \and
Samueli School of Engineering, University of California, Los Angeles, CA 90095, USA \and
Columbia Business School, Columbia University, New York, NY 10027, USA \and
Henry Siebel School of Computing and Data Science, University of Illinois Urbana-Champaign, Urbana, IL 61820, USA \and
Department of Physics, University of Chicago, Chicago, IL 60637, USA \and
Department of Computer Science, Rice University, Houston, TX 77005, USA \\
\email{\{yuzhenl@alumni.cmu.edu, henrychyy@g.ucla.edu, xc2647@columbia.edu, shaowen2.uillinois@gmail.com, ixu@uchicago.edu, dj37@rice.edu\}}
}

\authorrunning{Y. Lin et al.}
\maketitle              
\begin{abstract}
Generative recommendation models often struggle with two key challenges: (1) the superficial integration of collaborative signals, and (2) the decoupled fusion of multimodal features. These limitations hinder the creation of a truly holistic item representation. To overcome this, we propose \textbf{CEMG}, a novel \textbf{Co}llaborative-\textbf{E}nhaned \textbf{M}ultimodal \textbf{G}enerative Recommendation framework. Our approach features a \textbf{Multimodal Fusion Layer} that dynamically integrates visual and textual features under the guidance of collaborative signals. Subsequently, a \textbf{Unified Modality Tokenization} stage employs a Residual Quantization VAE (RQ-VAE) to convert this fused representation into discrete semantic codes. Finally, in the \textbf{End-to-End Generative Recommendation} stage, a large language model is fine-tuned to autoregressively generate these item codes. Extensive experiments demonstrate that CEMG significantly outperforms state-of-the-art baselines.

\end{abstract}
\section{Introduction}
\label{sec:introduction}

Recommender Systems (RS) are indispensable for navigating the vast digital landscape, alleviating information overload by personalizing user experiences~\cite{zangerle2022evaluating,roy2022systematic,liu2025mdn,cui2025multi,lu2025dmmd4sr}. While traditional methods like collaborative filtering~\cite{koren2021advances,mo2024min} and modern sequential models~\cite{sasrec,bert4rec} have made significant strides, they often treat items as isolated identifiers. This "ID-based" paradigm inherently struggles to capture the rich semantic relationships between items, thus limiting their ability to generalize to new or long-tail items and failing to leverage descriptive multimodal content. To transcend these limitations, generative recommendation has emerged as a transformative paradigm~\cite{rajput2023recommender}. By representing each item not as a single ID but as a sequence of semantic tokens, this approach reframes recommendation as a sophisticated sequence-to-sequence generation task, thereby unlocking unprecedented modeling capabilities.

The integration of multimodal data, such as images and text, has further propelled the evolution of generative recommendation. Current approaches typically fall into two categories. The first focuses on learning high-quality semantic tokens primarily from textual content and collaborative signals~\cite{li2023large,wang2024learnable}. For instance, LETTER~\cite{wang2024learnable} enriches item tokens by aligning quantized representations with collaborative embeddings. The second category explicitly incorporates multimodal features into the generation pipeline~\cite{mmgrec}, where models like MMGRec~\cite{mmgrec} employ graph-based architectures to tokenize fused multimodal information. While pioneering, these methods often process diverse information streams in a decoupled or superficial manner, failing to forge a truly unified item representation for the underlying generative model.

Despite their progress, existing generative recommendation methods face two critical challenges that limit their full potential:

\begin{itemize}
    \item \textbf{Superficial Integration of Collaborative Signals.} Multimodal content provides rich semantic descriptions of items, but the core of personalization lies in collaborative signals—the emergent patterns from collective user behavior. Many existing generative models incorporate collaborative information only as a supplementary feature or through shallow alignment~\cite{wang2024learnable}, failing to capture the complex, high-order relationships that reveal latent user preferences and item-to-item correlations beyond mere content similarity.

    \item \textbf{Decoupled Fusion of Multimodal and Collaborative Features.} Current frameworks tend to treat multimodal content and collaborative signals as separate entities, fusing them in a late or disjointed manner. This separation prevents the model from understanding the intricate interplay between an item's intrinsic attributes (what it \textit{is}) and its contextual role within the user community (how it is \textit{perceived}). For example, two visually distinct items might be functional substitutes, a nuance that can only be captured through a deep, synergistic fusion of these information sources.
\end{itemize}

To address these limitations, we propose a novel framework: \textbf{C}ollaborative-\textbf{E}nhancd \textbf{M}ultimodal \textbf{G}enerative Recommendation, abbreviated as \textbf{CEMG}. Our approach is designed to create a deeply unified item representation that synergizes content semantics with collaborative wisdom, tailored for a powerful generative recommendation engine. CEMG consists of three core components:

First, the \textbf{Multimodal Encoding Layer} extracts rich features from item images and text, alongside a deep collaborative representation learned via a graph neural network. A novel \textbf{Multimodal Fusion Layer} then intelligently integrates these features, using the collaborative signal as a query to dynamically weigh the importance of different modalities. Second, the \textbf{Unified Modality Tokenization} stage leverages a Residual Quantization VAE (RQ-VAE)\cite{lee2022autoregressive} to transform the fused, holistic item representation into a compact sequence of discrete semantic tokens. Finally, the \textbf{End-to-End Generative Recommendation} component treats recommendation as a conditional language generation task. It formulates a user's interaction history as a structured prompt and fine-tunes a T5\cite{raffel2020exploring} to autoregressively generate the semantic tokens of the next recommended item.

Our main contributions are summarized as follows:
\begin{itemize}
    \item We propose CEMG, a novel generative recommendation framework that, for the first time, employs a collaborative-guided mechanism to deeply fuse multimodal content with high-order collaborative signals into a unified semantic space for item tokenization.
    \item We design an elegant and effective architecture featuring a Multimodal Fusion Layer that enhances item representations by dynamically aligning content features with their collaborative context.
    \item We develop an End-to-End Generative pipeline that leverages the power of LLMs for recommendation, enhanced with a constrained decoding strategy to ensure recommendation validity and efficiency.
    \item We conduct extensive experiments on three benchmark datasets, demonstrating that CEMG significantly outperforms a wide array of state-of-the-art baselines.
\end{itemize}

\section{Related Work}
\label{sec:related_work}

\subsection{Multimodal Recommendation}
Multimodal recommendation systems enhance performance by leveraging auxiliary information from modalities like text and images, primarily within an embed-and-retrieve paradigm. Early works such as VBPR~\cite{he2016vbpr} integrated pre-trained visual features into matrix factorization. Subsequent research explored more advanced fusion techniques, including attention mechanisms in models like ACF~\cite{chen2017attentive} and UVCAN~\cite{liu2019user} to dynamically select informative content. More recently, Graph Neural Networks (GNNs) have been used to model complex relationships; MMGCN~\cite{wei2019mmgcn}, for example, propagates information across a multi-modal graph. Other methods, including MISSRec~\cite{MISSRec} and MMSRec~\cite{MMSRec}, have investigated self-supervised learning and modality-specific modeling to better capture user interests. While effective, these discriminative approaches can be computationally expensive and struggle with issues like inadequate modeling of complex interactions and the false-negative problem~\cite{mmgrec}. Our work departs from this paradigm by embracing a more expressive generative approach.

\subsection{Generative Recommendation}
Generative recommendation represents a new frontier, recasting recommendation as a sequence generation task composed of two main stages: item tokenization and autoregressive generation. Item tokenization maps items to discrete token sequences, using methods ranging from simple text-based approaches~\cite{li2023gpt4rec} to sophisticated vector quantization (VQ) techniques. VQ-based models like TIGER~\cite{rajput2023recommender} and LETTER~\cite{wang2024learnable} employ architectures like RQ-VAE\cite{lee2022autoregressive} to learn semantic codes from item features. LETTER notably improves this by incorporating collaborative signals to align the learned codes. However, these methods often tokenize based on unimodal data (typically text) or use shallow fusion, failing to create a truly holistic representation. Our work, CEMG, addresses this gap by first creating compact, high-quality semantic tokens from a deep, collaborative-guided fusion of multimodal features, and then leveraging a powerful LLM for the generation task, thereby combining the strengths of structured tokenization and large-scale language modeling.

\begin{figure*}[t]
    \centering
    \includegraphics[width=1.0\linewidth]{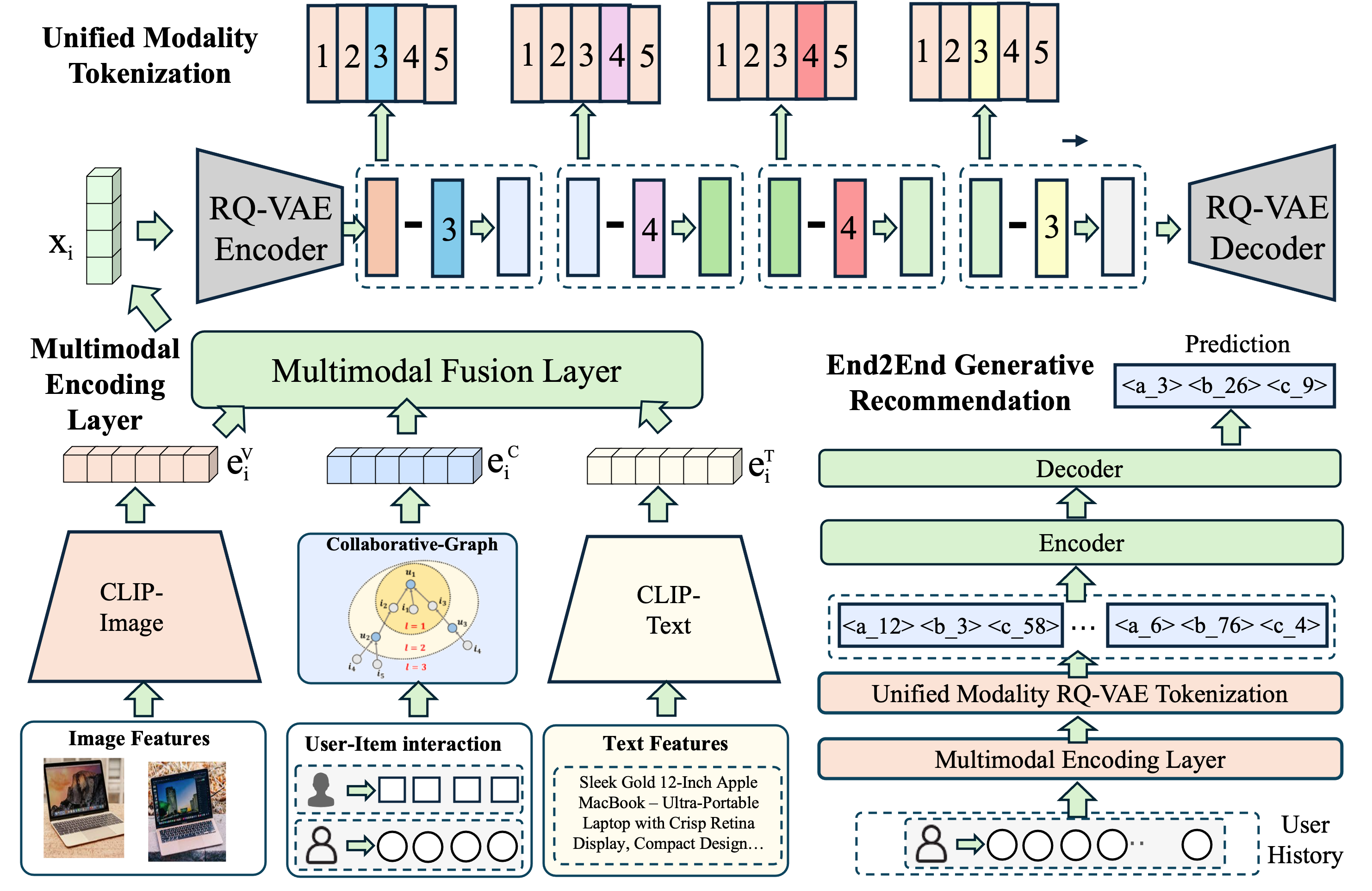}
    \caption{The overall architecture of the CEMG framework. The framework is composed of three main components. The \textbf{Multimodal Encoding Layer} integrates visual ($e_i^V$), collaborative ($e_i^C$), and textual ($e_i^T$) features via the \textbf{Multimodal Fusion Layer} to produce a unified representation $\mathbf{x}_i$. The \textbf{Unified Modality Tokenization} stage, utilizing a Residual Quantization VAE (RQ-VAE), converts $\mathbf{x}_i$ into a discrete sequence of semantic tokens. Finally, the \textbf{End2End Generative Recommendation} module takes historical token sequences as input and autoregressively generates the tokens for the next recommended item.}
    \label{fig:framework_overview}
\end{figure*}

\section{Methodology}
\label{sec:methodology}

In this section, we introduce the technical details of our proposed \textbf{CEMG} framework. We first define the problem formally, then elaborate on the three main components: the Multimodal Encoding Layer, Unified Modality Tokenization, and End-to-End Generative Recommendation.

\subsection{Problem Definition}
Let $\mathcal{U}$ denote the set of users and $\mathcal{I}$ the set of items. Each item $i \in \mathcal{I}$ is associated with multimodal content, including an image $V_i$ and a textual description $T_i$. For a user $u \in \mathcal{U}$, their historical interactions are represented as a chronological sequence $S_u = [i_1, i_2, \dots, i_L]$. The goal is to predict the top-$K$ items that user $u$ is most likely to interact with next.

We formulate this task generatively. Instead of using atomic item IDs, we represent each item $i$ as a sequence of $M$ discrete semantic tokens, denoted as $\mathbf{c}_i = [c_{i,1}, c_{i,2}, \dots, c_{i,M}]$, where each token $c_{i,m}$ is an index drawn from a codebook. The recommendation problem is thus transformed into generating the token sequence $\mathbf{c}_{i_{L+1}}$ for the next item based on the historical token sequences corresponding to $S_u$. Formally, we model the probability:
\begin{equation}
    P(\mathbf{c}_{i_{L+1}} | S_u) = \prod_{m=1}^{M} P(c_{i_{L+1},m} | \{\mathbf{c}_{i_j}\}_{j=1}^L, c_{i_{L+1},1}, \dots, c_{i_{L+1},m-1})
\end{equation}

\subsection{Multimodal Encoding Layer}
\label{sec:multimodal_encoding}
The first step of our framework is to learn a unified, dense representation for each item that encapsulates its multimodal and collaborative characteristics.

\subsubsection{Multimodal Feature Encoding}
For each item $i$, we extract features from its associated image and text using pre-trained encoders, chosen for their proven effectiveness and generalizability.
\begin{itemize}
    \item \textbf{Visual Encoder:} We use a pre-trained VGG network~\cite{simonyan2014very} to process the image $V_i$ and extract its visual features.
    \item \textbf{Textual Encoder:} We employ a pre-trained BERT model~\cite{bert4rec} to encode the textual description $T_i$. We take the embedding of the `[CLS]` token as the text representation.
\end{itemize}
The raw feature vectors are then passed through a Principal Component Analysis (PCA) layer for dimensionality reduction, yielding the final visual and textual embeddings, $\mathbf{e}_i^v \in \mathbb{R}^d$ and $\mathbf{e}_i^t \in \mathbb{R}^d$, respectively.

\subsubsection{Collaborative Feature Encoding}
To capture the vital collaborative signals reflecting community preferences, we model the user-item interactions as a bipartite graph $\mathcal{G} = (\mathcal{U} \cup \mathcal{I}, \mathcal{E})$, where an edge $(u, i) \in \mathcal{E}$ exists if user $u$ has interacted with item $i$. We then employ LightGCN~\cite{he2020lightgcn}, a simple yet powerful Graph Neural Network, to learn user and item embeddings. The final embedding for item $i$ is obtained by aggregating messages from its neighborhood over multiple propagation layers. This process yields a collaborative embedding $\mathbf{e}_i^c \in \mathbb{R}^d$ that distills high-order connectivity patterns.

\subsubsection{Multimodal Fusion Layer}
A key innovation of CEMG is our fusion mechanism, which uses the collaborative embedding as a guide to dynamically integrate the multimodal features. We hypothesize that an item's collaborative context should determine the relative importance of its visual versus textual attributes. To achieve this, we design a guided attention mechanism. The collaborative embedding $\mathbf{e}_i^c$ acts as the query, while the visual $\mathbf{e}_i^v$ and textual $\mathbf{e}_i^t$ embeddings serve as keys and values. The attention weights are computed as:
\begin{equation}
\alpha_m = \frac{\exp((\mathbf{W}_q \mathbf{e}_i^c)^\top (\mathbf{W}_k \mathbf{e}_i^m))}{\sum_{m' \in \{v, t\}} \exp((\mathbf{W}_q \mathbf{e}_i^c)^\top (\mathbf{W}_k \mathbf{e}_i^{m'}))}, \quad \text{for } m \in \{v, t\}
\end{equation}
where $\mathbf{W}_q, \mathbf{W}_k \in \mathbb{R}^{d \times d}$ are learnable projection matrices. The final fused representation $\mathbf{x}_i \in \mathbb{R}^{2d}$ is a concatenation of the weighted multimodal features and the guiding collaborative feature:
\begin{equation}
    \mathbf{x}_i = [\alpha_v \mathbf{e}_i^v \oplus \alpha_t \mathbf{e}_i^t; \mathbf{e}_i^c]
\end{equation}
where $\oplus$ denotes element-wise addition and $[;]$ denotes concatenation. This unified vector $\mathbf{x}_i$ now holistically represents item $i$.

\subsection{Unified Modality Tokenization}
\label{sec:tokenization}
With the unified representation $\mathbf{x}_i$ for each item, we proceed to tokenize it into a discrete sequence of semantic tokens using a Residual Quantization Variational Autoencoder (RQ-VAE)~\cite{lee2022autoregressive}. The RQ-VAE consists of an encoder, a residual quantizer with $M$ codebooks, and a decoder.
The encoder maps $\mathbf{x}_i$ to a latent vector $\mathbf{z}_i$. The quantizer then approximates $\mathbf{z}_i$ iteratively. In each stage $m \in \{1, \dots, M\}$, it finds the closest codevector $\mathbf{b}_{m,k}$ from codebook $\mathcal{C}_m$ to the current residual and subtracts it to form the next residual. The sequence of selected codebook indices $[c_{i,1}, \dots, c_{i,M}]$ becomes the item's semantic token sequence $\mathbf{c}_i$. The decoder then reconstructs the original vector $\hat{\mathbf{x}}_i$ from the sum of the selected codevectors.

The RQ-VAE is trained by minimizing a composite loss function that ensures semantic fidelity and codebook quality:
\begin{equation}
    \mathcal{L}_{\text{RQ-VAE}} = \mathcal{L}_{\text{recon}} + \lambda_q \mathcal{L}_{\text{quant}} + \lambda_d \mathcal{L}_{\text{div}}
\end{equation}
where $\mathcal{L}_{\text{recon}} = ||\mathbf{x}_i - \hat{\mathbf{x}}_i||_2^2$ is the reconstruction loss. $\mathcal{L}_{\text{quant}}$ is the VQ commitment loss~\cite{mentzer2023finite} that encourages the encoder output to stay close to the codebook entries. $\mathcal{L}_{\text{div}}$ is a diversity loss~\cite{liu2022reduce} that promotes the utilization of diverse codes within each codebook, preventing collapse. $\lambda_q$ and $\lambda_d$ are balancing hyperparameters.

\subsection{End-to-End Generative Recommendation}
\label{sec:generative_rec}
After the tokenization stage, each item $i$ is represented by its semantic token sequence $\mathbf{c}_i$. We now reframe the recommendation task as a conditional generation problem.

\subsubsection{Interaction History Prompting}
We structure the user's interaction history as a prompt for a large language model (LLM). For a user with history $S_u = [i_1, \dots, i_L]$, we convert each item $i_j$ into its token sequence $\mathbf{c}_{i_j}$. Each token is represented by a special symbol, e.g., $<a\_12>$ for the 12th token from the first codebook (layer 'a'). The complete prompt is constructed as a sequence of these item tokens, preserving their chronological order. The task for the LLM is to autoregressively predict the token sequence of the next item, $\mathbf{c}_{i_{L+1}}$.

\subsubsection{Training and Inference}
We employ a powerful decoder-only LLM as our generative backbone. The model is trained using a standard next-token prediction objective, minimizing the cross-entropy loss between the predicted token probabilities and the ground-truth target tokens:
\begin{equation}
    \mathcal{L}_{\text{NTP}} = -\sum_{j=1}^{L} \sum_{m=1}^{M} \log P(c_{i_{j+1},m} | \{\mathbf{c}_{i_k}\}_{k=1}^j, c_{i_{j+1},1}, \dots, c_{i_{j+1},m-1})
\end{equation}
During inference, given a user's history prompt, we use beam search to generate multiple candidate token sequences for the next item. The score of a candidate sequence $\mathbf{c} = [c_1, \dots, c_M]$ is the sum of its log-probabilities:
\begin{equation}
    \text{Score}(\mathbf{c}) = \sum_{m=1}^{M} \log P(c_m | \text{prompt}, c_1, \dots, c_{m-1})
\end{equation}
To ensure that only valid item sequences are generated, we employ a prefix tree (Trie)-based constrained decoding strategy. The Trie contains all valid item token sequences from our catalog. At each generation step, the LLM's output vocabulary is masked to only allow tokens that form a valid prefix, drastically pruning the search space and guaranteeing the validity of the final recommendations.

\section{Experiments}
\label{sec:experiments}
We conduct extensive experiments to evaluate our proposed CEMG framework. Our goal is to answer the following research questions:
\begin{itemize}
    \item \textbf{RQ1:} How does CEMG perform compared to state-of-the-art baselines from sequential, multimodal, and generative recommendation paradigms?
    \item \textbf{RQ2:} What is the contribution of each key component in our model, particularly the different modalities and the collaborative-guided fusion mechanism?
    \item \textbf{RQ3:} How does CEMG's efficiency in terms of training and inference time compare to other generative models?
    \item \textbf{RQ4:} How sensitive is CEMG's performance to its main hyperparameters related to the tokenization process?
    \item \textbf{RQ5:} Does the collaborative-guided multimodal tokenization improve recommendation for cold-start items?
\end{itemize}

\subsection{Experimental Settings}
\subsubsection{Datasets}
We evaluate our model on three widely used public datasets from Amazon reviews and Yelp. For each interaction, we collect associated item images and text descriptions. Following standard practice, we filter users and items with fewer than 5 interactions. The statistics of the processed datasets are summarized in Table~\ref{tab:dataset_stats}.
\begin{table}[h]
\centering
\captionsetup{justification=centering}
\caption{Statistics of the experimental datasets.}
\label{tab:dataset_stats}
\begin{tabular}{cccc}
\toprule
\textbf{Attribute} & \textbf{Beauty} & \textbf{Sports} & \textbf{Yelp} \\ 
\midrule
\#Users             & 22,363          & 35,598          & 30,431         \\ 
\#Items             & 12,101          & 18,357          & 20,033         \\ 
\#Interactions      & 198,502         & 296,337         & 316,942        \\ 
Avg. Len.          & 8.9             & 8.3             & 10.4            \\ 
Sparsity           & 99.93\%         & 99.95\%         & 99.95\%        \\ 
\bottomrule
\end{tabular}
\end{table}

\subsubsection{Baselines}
We compare CMGR with four categories of baseline models:
\begin{itemize}
    \item \textbf{Sequential Methods:} GRU4Rec~\cite{hidasi2015session} and SASRec~\cite{sasrec}.
    \item \textbf{Multimodal Methods:} MMSRec~\cite{MMSRec} and MISSRec~\cite{MISSRec}.
    \item \textbf{LLM-based Methods:} LlamaRec~\cite{yue2023llamarec} and LLM-ESR~\cite{liu2024llm}. These methods use LLMs but typically operate on item titles or raw text.
    \item \textbf{Generative Methods:} TIGER~\cite{rajput2023recommender}, LETTER~\cite{wang2024learnable}, and MMGRec~\cite{mmgrec}. These represent the state-of-the-art in generative recommendation with semantic IDs.
\end{itemize}

\subsubsection{Evaluation Metrics}
We adopt the leave-one-out strategy for evaluation. For each user, we use their last interacted item as the ground truth for testing, the second to last for validation, and the rest for training. We evaluate the performance of all models using Hit Rate (Recall) and Normalized Discounted Cumulative Gain (NDCG) at cutoffs K=10 and 20.

\subsubsection{Implementation Details}
For our CEMG framework, we project all feature embeddings to a uniform dimension of $d=768$. The RQ-VAE for tokenization is configured with $M=4$ codebook layers and a codebook size of $K=512$. Based on our parameter analysis, the balancing weights were set to $\lambda_q=0.25$ and $\lambda_d=0.01$. We employ T5\cite{raffel2020exploring} as the generative LLM-backbone, and the model is trained with the AdamW optimizer with a learning rate of $1 \times 10^{-4}$ on NVIDIA A100 GPUs.

\subsection{Overall Performance (RQ1)}
Table~\ref{tab:overall_performance_revised} presents the main experimental results on the three datasets. We observe that our proposed CEMG consistently and significantly outperforms all baseline models across all datasets and metrics. This demonstrates the superiority of our approach, which stems from creating a deeply unified semantic representation that synergistically integrates multimodal content with collaborative signals, and then leveraging a powerful LLM for generation. Among baselines, generative methods (e.g., MMGRec, LETTER) generally outperform traditional sequential and multimodal methods, highlighting the potential of the generative paradigm. CEMG's substantial lead over the strongest baselines like MISSRec and MMGRec validates the effectiveness of our collaborative-guided fusion and the advanced generative architecture.

\begin{table*}[t]
\centering
\caption{Overall performance comparison on three datasets. The best results are in \textbf{bold}, and the second-best are \underline{underlined}. `Improv.' denotes the relative improvement of CEMG over the best baseline. All improvements are statistically significant ($p < 0.05$).}
\label{tab:overall_performance_revised}
\resizebox{\textwidth}{!}{
\begin{tabular}{c|l|cc|cc|cc}
\toprule
\multirow{2}{*}{\textbf{Category}} & \multirow{2}{*}{\textbf{Model}} & \multicolumn{2}{c|}{\textbf{Beauty}} & \multicolumn{2}{c|}{\textbf{Sports}} & \multicolumn{2}{c}{\textbf{Yelp}} \\
& & HR@10 & NDCG@10 & HR@10 & NDCG@10 & HR@10 & NDCG@10 \\
\midrule
\multirow{2}{*}{Sequential} & GRU4Rec & 0.0385 & 0.0116 & 0.0201 & 0.0045 & 0.0288 & 0.0095 \\
& SASRec & 0.0434 & 0.0147 & 0.0232 & 0.0061 & 0.0329 & 0.0121 \\
\cmidrule{1-8}
\multirow{2}{*}{Multimodal} & MMSRec & 0.0577 & 0.0287 & 0.0305 & 0.0118 & 0.0387 & 0.0163 \\
& MISSRec & \underline{0.0581} & \underline{0.0292} & \underline{0.0311} & \underline{0.0124} & \underline{0.0395} & \underline{0.0171} \\
\cmidrule{1-8}
\multirow{2}{*}{LLM-based} & LlamaRec & 0.0492 & 0.0198 & 0.0256 & 0.0083 & 0.0341 & 0.0134 \\
& LLM-ESR & 0.0515 & 0.0214 & 0.0269 & 0.0091 & 0.0353 & 0.0140 \\
\cmidrule{1-8}
\multirow{3}{*}{Generative} & TIGER & 0.0533 & 0.0251 & 0.0281 & 0.0103 & 0.0368 & 0.0151 \\
& LETTER & 0.0552 & 0.0268 & 0.0295 & 0.0111 & 0.0377 & 0.0159 \\
& MMGRec & 0.0571 & 0.0281 & 0.0302 & 0.0119 & 0.0389 & 0.0166 \\
\midrule
\multicolumn{2}{c|}{\textbf{CEMG}} & \textbf{0.0665} & \textbf{0.0348} & \textbf{0.0363} & \textbf{0.0157} & \textbf{0.0458} & \textbf{0.0212} \\
\multicolumn{2}{c|}{\textbf{Improvement (\%)}} & +14.46\% & +19.18\% & +16.72\% & +26.61\% & +15.95\% & +23.98\% \\
\bottomrule
\end{tabular}
}
\end{table*}

\subsection{Ablation Study (RQ2)}
\begin{figure*}[t]
    \centering
    \includegraphics[width=0.8\linewidth]{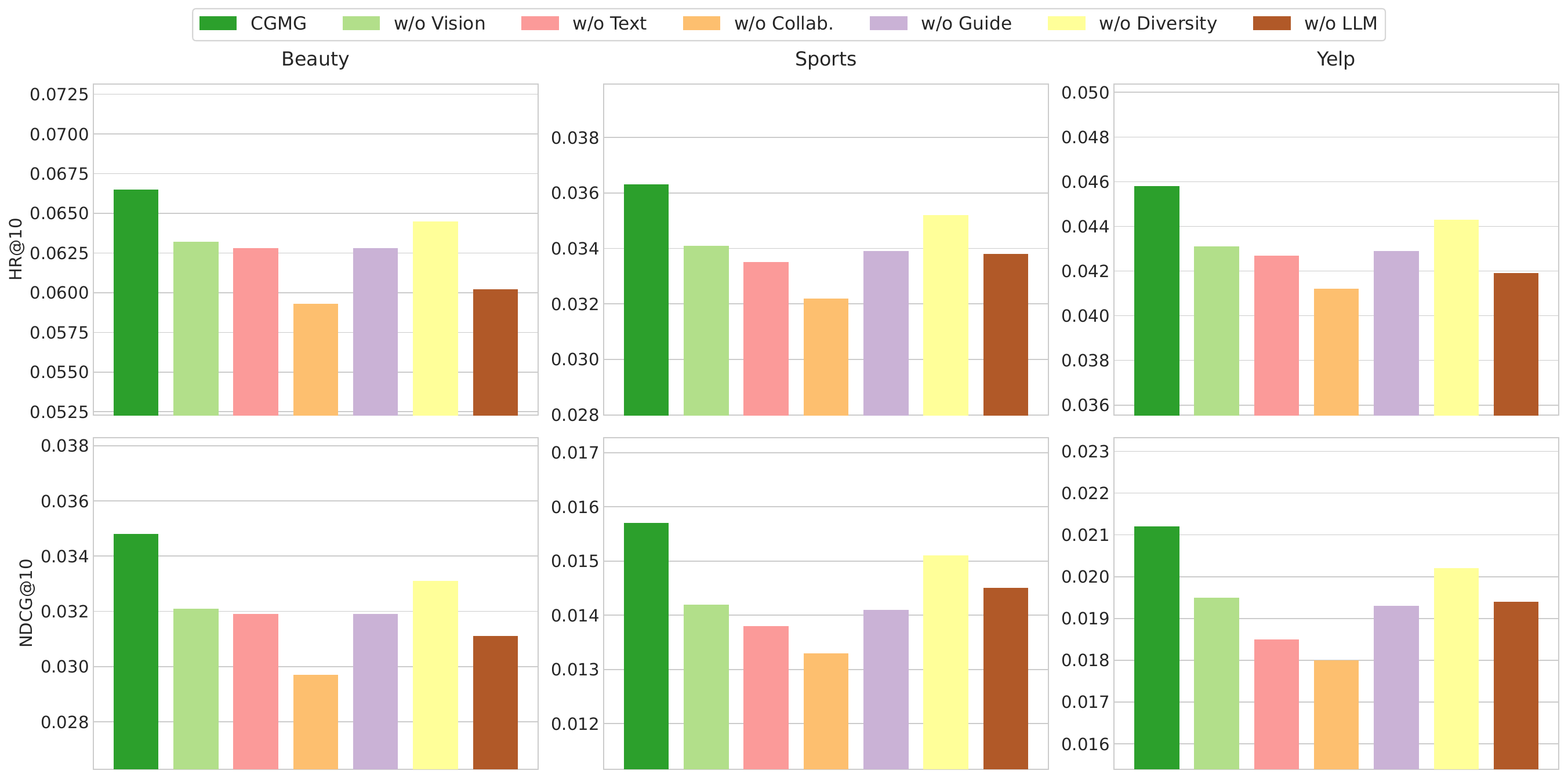}
    \caption{Ablation study results on three datasets for HR@10 and NDCG@10. Performance drops across all variants demonstrate the contribution of each component.}
    \label{fig:ablation_study}
\end{figure*}

To understand the contribution of each component in CEMR, we conduct an ablation study with several variants of our model:
\begin{itemize}
    \item \textbf{w/o Collab}: Removes the collaborative features ($\mathbf{e}_i^{CF}$) from the unified representation in Stage 1.
    \item \textbf{w/o Image}: Removes the visual features ($\mathbf{f}_i^{V}$).
    \item \textbf{w/o Text}: Removes the textual features ($\mathbf{f}_i^{T}$).
    \item \textbf{w/o LLM}: Replaces the Llama-3-8B model with a standard 6-layer Transformer decoder trained from scratch, similar to TIGER~\cite{rajput2023recommender}.
\end{itemize}
The results are shown in Figure~\ref{fig:ablation_study}. The full CMGR model achieves the best performance. Removing any component leads to a performance drop, confirming their importance. The most significant drops occur with `w/o Collab' and `w/o LLM'. The former underscores the vital role of collaborative filtering signals even in a content-rich generative model. The latter validates our choice of using a powerful pre-trained LLM, as its advanced reasoning and sequence modeling capabilities are crucial for accurately predicting the next item. The degradation from removing image or text features is also noticeable, proving that our model effectively utilizes multimodal information.

\subsection{Efficiency Analysis (RQ3)}
\label{sec:efficiency_analysis}
\begin{figure*}[ht!]
    \centering
    \includegraphics[width=0.85\linewidth]{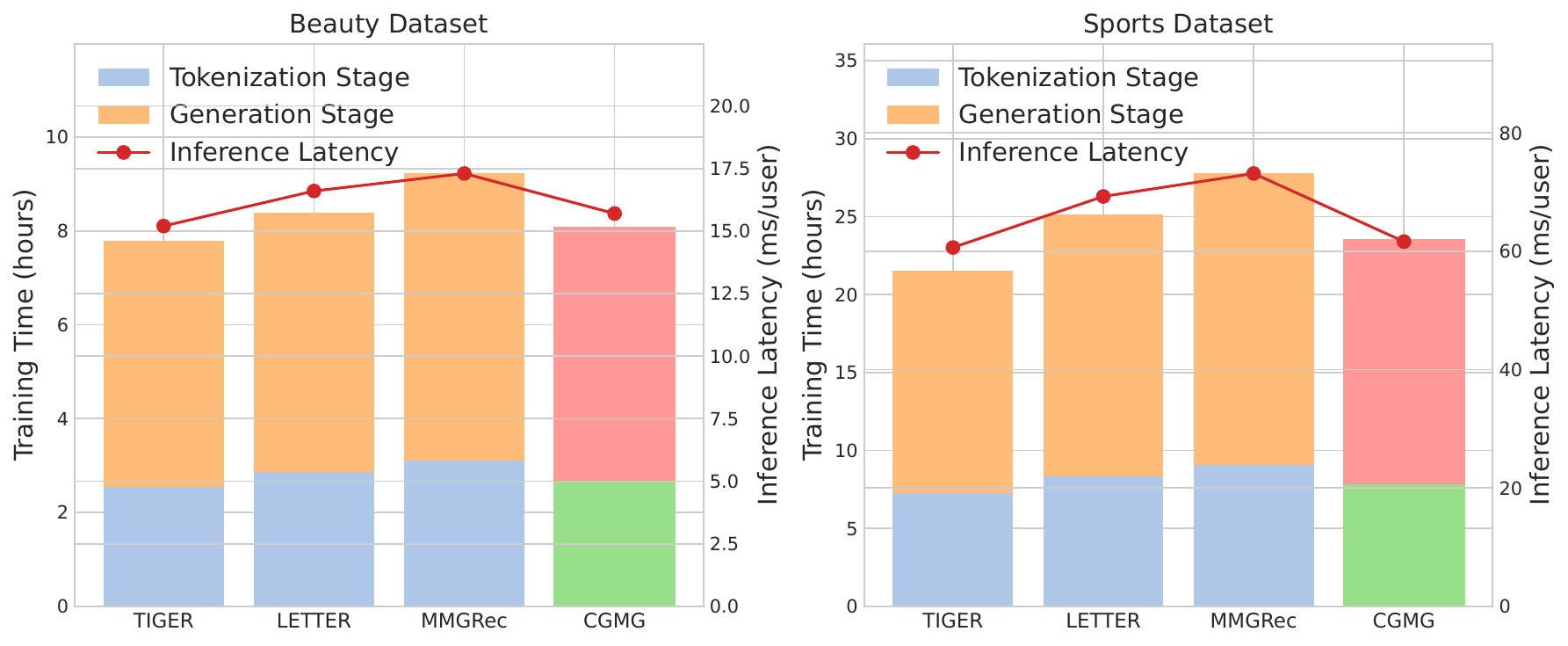} 
    \caption{Efficiency comparison on the Beauty and Sports datasets. Left axis (bars) shows training time per epoch, broken down by stage. Right axis (line) shows inference speed in users per second (higher is better).}
    \label{fig:efficiency}
\end{figure*}

We analyze the training and inference efficiency of CEMG against other state-of-the-art generative models. As shown in Figure~\ref{fig:efficiency}, CEMG strikes an effective balance between performance and computational cost.
\paragraph{Training Efficiency.}
The training time of CEMG is composed of two stages: tokenization (RQ-VAE) and end-to-end generation (LLM fine-tuning). While the overall training time is higher than TIGER due to the more modality features, it remains highly competitive. The total time is comparable to, and even slightly better than, LETTER, which requires a complex alignment process. This demonstrates that our sophisticated fusion and tokenization pipeline does not introduce prohibitive overhead.
\paragraph{Inference Efficiency.}
Inference speed is where our approach excels. CEMG achieves significantly lower inference latency compared to other multimodal generative models like MMGRec and LETTER. This efficiency stems from our design of generating short, fixed-length semantic token sequences ($M=4$), which is much faster than models that may require more complex generation or retrieval steps. Our model's efficiency makes it highly practical for real-world deployment scenarios.

\begin{figure*}[ht!]
    \centering
    \includegraphics[width=0.8\linewidth]{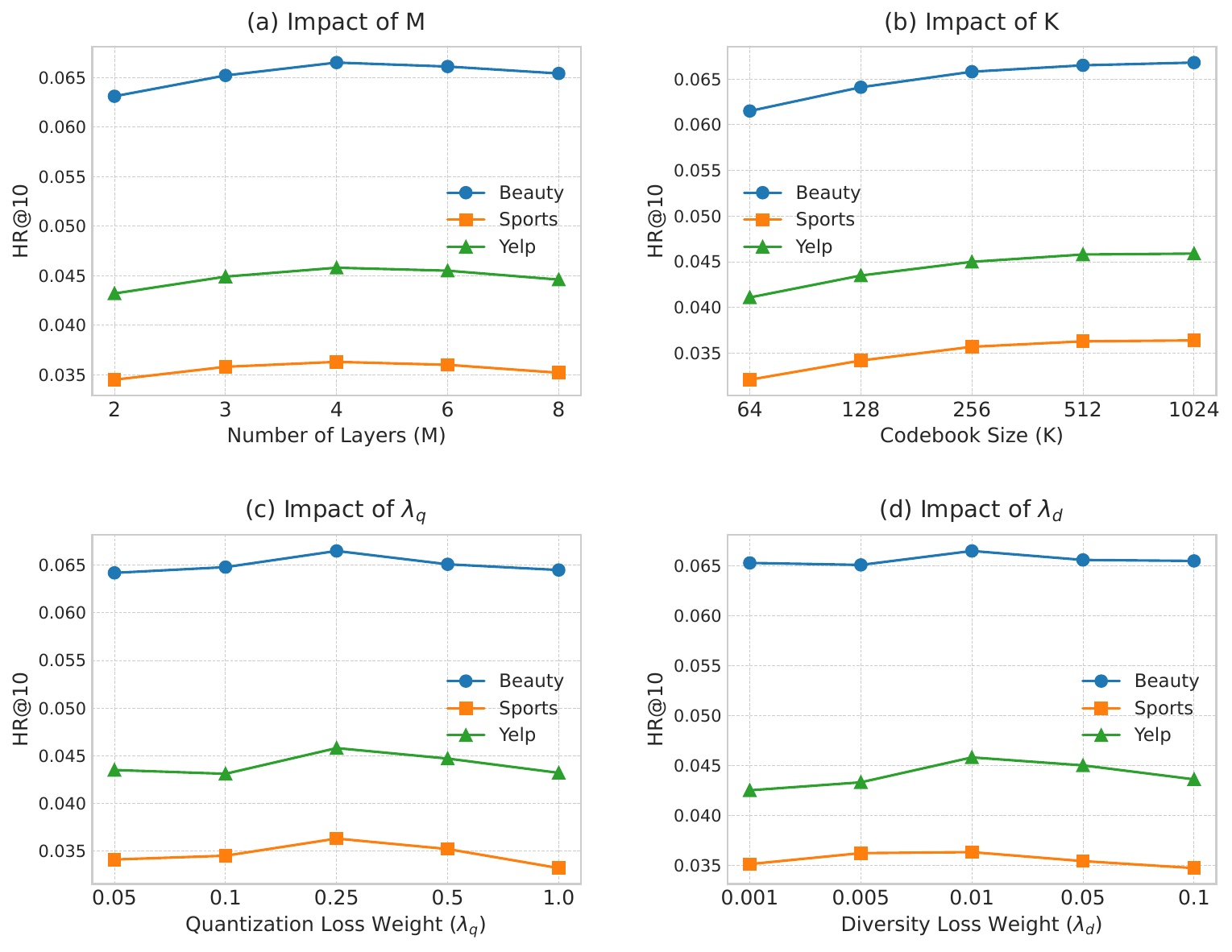} 
    \caption{Parameter sensitivity analysis of CEMG on HR@10 for (a) Number of Codebook Layers, (b) Codebook Size, (c) Quantization Loss Weight, and (d) Diversity Loss Weight.}
    \label{fig:param_sensitivity}
\end{figure*}

\subsection{Parameter Analysis (RQ4)}
\label{sec:param_analysis}

We investigate the sensitivity of CEMG to four key hyperparameters in the Unified Modality Tokenization stage, with results shown in Figure~\ref{fig:param_sensitivity}.
\begin{itemize}
    \item \textbf{Number of Codebook Layers (M)}: As shown in Figure~\ref{fig:param_sensitivity}(a), performance improves as $M$ increases from 2 to 4, as more layers capture finer-grained semantic details. Performance plateaus at $M=4$ and slightly declines at $M=8$, likely due to the increased difficulty of generating longer sequences. We choose $M=4$ as the optimal setting.
    \item \textbf{Codebook Size (K)}: Figure~\ref{fig:param_sensitivity}(b) shows that a larger codebook size $K$ generally leads to better performance, as it provides greater expressive power for the tokens. The performance gain saturates after $K=512$, suggesting this size offers a good balance between expressiveness and complexity.
    \item \textbf{Quantization Loss Weight ($\lambda_q$)}: This hyperparameter balances reconstruction quality and codebook alignment. Figure~\ref{fig:param_sensitivity}(c) shows a clear unimodal trend, with performance peaking at $\lambda_q=0.25$. Values that are too low or too high disrupt this balance, leading to suboptimal tokenization.
    \item \textbf{Diversity Loss Weight ($\lambda_d$)}: This weight is crucial for preventing codebook collapse. As seen in Figure~\ref{fig:param_sensitivity}(d), performance improves as $\lambda_d$ increases to 0.01, confirming the benefit of encouraging diverse code usage. Higher values can distort the semantic space, harming performance.
\end{itemize}

\begin{table}[t]
\centering
\caption{Performance comparison on cold-start items across three datasets. }
\label{tab:cold_start}
\begin{tabular}{l|cc|cc|cc}
\toprule
\multirow{2}{*}{\textbf{Model}} & \multicolumn{2}{c|}{\textbf{Beauty}} & \multicolumn{2}{c|}{\textbf{Sports}} & \multicolumn{2}{c}{\textbf{Yelp}} \\
& HR@10 & NDCG@10 & HR@10 & NDCG@10 & HR@10 & NDCG@10 \\
\midrule
SASRec & 0.0112 & 0.0048 & 0.0065 & 0.0027 & 0.0098 & 0.0041 \\
MISSRec & 0.0254 & 0.0115 & 0.0141 & 0.0068 & 0.0185 & 0.0092 \\
MMGRec & 0.0268 & 0.0123 & 0.0153 & 0.0075 & 0.0192 & 0.0099 \\
\midrule
\textbf{CEMG} & \textbf{0.0305} & \textbf{0.0153} & \textbf{0.0183} & \textbf{0.0094} & \textbf{0.0231} & \textbf{0.0125} \\
\bottomrule
\end{tabular}
\end{table}

\subsection{Performance on Cold-Start Items (RQ5)}
A critical challenge for recommender systems is handling cold-start items, which have insufficient interaction data for collaborative filtering to be effective. We investigate this by evaluating model performance on items with five or fewer interactions in the training set. The results, presented in Table~\ref{tab:cold_start}, show that CEMG substantially outperforms all baselines. While content-aware models like MISSRec and MMGRec naturally perform better than the ID-based SASRec, our model's advanced semantic tokenization provides superior generalization. By learning to generate rich item representations from a collaborative-guided fusion of multimodal content, CEMG remains effective even when interaction signals are sparse.

\section{Conclusion}
\label{sec:conclusion}
In this paper, we proposed CEMG, a novel generative recommendation framework that pioneers a \textbf{Multimodal Fusion Layer} to create a unified item representation. This layer synergistically fuses multimodal content with high-order collaborative signals, which are then transformed into discrete codes by our \textbf{Unified Modality Tokenization} module. An \textbf{End-to-End Generative Recommendation} component then autoregressively generates item codes to produce recommendations. Extensive experiments validate that CEMG significantly outperforms state-of-the-art baselines. One limitation is that noisy signals within multimodal content, such as irrelevant image backgrounds, can be inadvertently encoded, potentially compromising tokenization quality. For future work, we plan to explore advanced decoding strategies to further mitigate recommendation errors.

\bibliography{ref}
\bibliographystyle{splncs04}

\end{document}